\documentclass[useAMS,usenatbib]{mn2e}
\usepackage{graphicx,rotate,url,mathptmx,lscape,times,amssymb,color}
\def\gtsim{\mathrel{\hbox{\rlap{\hbox{\lower4pt\hbox{$\sim$}}}\hbox{$>$}}}}
\def\lesssim{\mathrel{\hbox{\rlap{\hbox{\lower4pt\hbox{$\sim$}}}\hbox{$<$}}}}

\def\ga{{\rm\thinspace gauss}}

\def\h0{\hbox{{\rm H}$^0$}}
\DeclareMathAlphabet{\vib}{OML}{cmm}{m}{it}

\title[Corrigendum: Improved He~I Emissivities]{Corrigendum: Improved He~I Emissivities in the Case B Approximation}

\author[R.L. Porter et al.]
       {\parbox[]{6.0in}
        {R.L. Porter$^{1}$\thanks{E-mail: ryanlporter@gmail.com},
        G.J. Ferland$^{2}$,
        P.J. Storey$^{3}$,
        M.J. Detisch$^{2}$\\
        \footnotesize
        $^1$Department of Physics and Astronomy and Center for Simulational Physics, University of Georgia, Athens, GA 30602, USA\\
        $^2$Dept. of Physics \& Astronomy, University of Kentucky, Lexington, KY 40506, USA\\
        $^3$Dept. of Physics \& Astronomy, University College London, Gower Street, London WC1E 6BT, UK}}
\date{
      Received }

\pagerange{\pageref{firstpage}--\pageref{lastpage}}
\pubyear{}

\begin{document}

\maketitle

\label{firstpage}



\setcounter{figure}{2}
\setcounter{table}{1}

A setup error caused allowed resonance lines to escape via scattering from free electrons.
Transitions to the ground state should not escape in the Case-B approximation.
The escaping line photons resulted in decreased populations of $np\,^1\!P$ levels, 
and indirectly decreased populations of other levels (via radiative decays and collisions).
This most strongly affected low-$L$ singlet transitions at densities $\lesssim10^5$~cm$^{-3}$.

We have turned off the process and recalculated our results.
Corrections to lines emitted from $np\,^1\!P$ levels can be more than an order of magnitude,
while lines from $ns\,^1\!S$ levels are corrected by up to a factor of $\sim2$.  
This affected 11 of the 44 lines reported in the supplemental table.
Most lines are affected by $\sim1\%$ or less.
All line emissivities increase (or are negligibly affected) due to this change.

An additional error was the inadvertent disabling of some collisions with $\Delta n >5$.
This slowed approach to local thermodynamic equilibrium with increasing temperature or density,
but the effects are generally comparable to or less than the uncertainties due
to collisional rates.
This omission has also been corrected here.
Line emissivities can both increase and decrease as a result of this change.
The behavior is a function of temperature and density.

Figure~1 pertained only to fundamental data and not the results of simulations.
It is unaffected by the error.
Figures~2 and 4 are only weakly affected.  
The identified trends are unchanged and reproducing those figures is unnecessary. 

Of the six emissivity ratios in Figure~3, which is re-plotted here, four of them are only weakly affected.  
The results for $\lambda\lambda 5876$ and $6678$ have increased as a result of the changes described here,
the latter because its upper level, $3d\,^1\!D$, is strongly populated by radiative decays from higher $np\,^1\!P$ levels,
the former because $3d\,^1\!D$ and $3d\,^3\!D$ are strongly mixed collisionally.
These changes are in the same direction but smaller than the ones reported in the original manuscript. 

We also compared our new emissivities to the full set of Benjamin, Skillman, \& Smits (1999; hereafter BSS99)
results at $10,000$K and $n_e = 100$~cm$^{-3}$.
The largest difference ($\sim6\%$) is for $\lambda 17003$ and seems to be directly attributable
to different absorption oscillator strengths published by Kono \& Hattori (1984) and Drake (1996).
Only 12 of the remaining 32 emissivities differ by more than $1\%$ -- the largest by $\sim3\%$. 
The differences are strongly correlated with differences in recombination coefficients.
Much larger differences continue to exist at higher densities and temperatures.

Table~1 contained a line list and associated level designations and does not require corrections.
Table~2  and the supplemental table have been updated.  

\section*{Acknowledgments}

The authors thank Yuri Izotov and Manuel Peimbert for separately contacting us about the erroneous results.


%


\begin{figure}
\protect\resizebox{\columnwidth}{!}
{\includegraphics{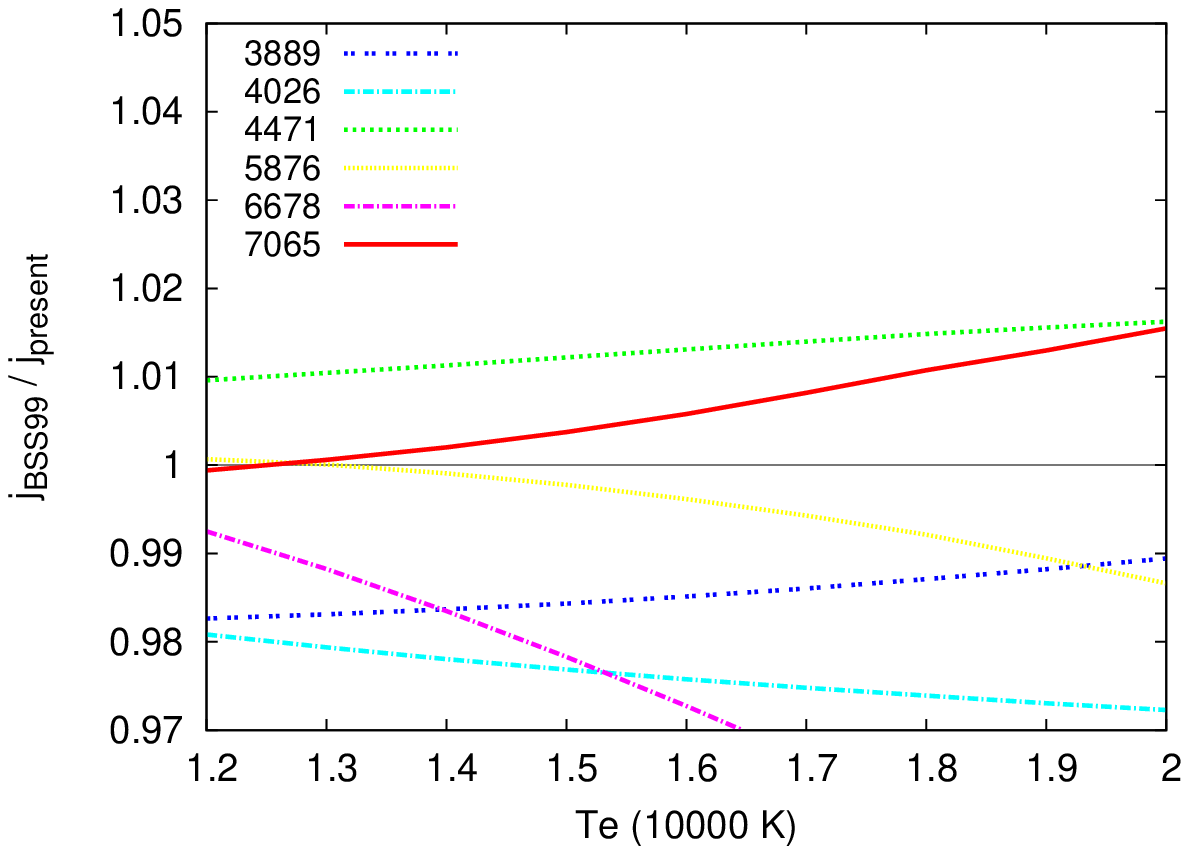}}
\caption{Ratio of BSS99 and present emissivities for several strong lines as a function of temperature with $n_e = 100$~cm$^{-3}$.  
}
\label{fig:BSS99}
\end{figure}


\begin{table*}
\centering
\caption{Emissivities of several He~I lines at conditions important for primordial abundance analyses.
This table is a small subset of the full results.  
Values are $4\pi j/n_e n_{\mathrm{He+}}$ in units $10^{-25}$~erg~cm$^{3}$~s$^{-1}$.}
\begin{tabular}{rrrrrrrr}
\hline
$T_e$~(K)& $n_e$ (cm$^{-3}$)&	 3889\AA	& 	4026\AA	&	 4471\AA	&	 5876\AA	&	 6678\AA	&	 7065\AA	\\
\hline
10000 & 10 	& 1.3897 & 0.2905 & 0.6105 & 1.6838 & 0.4788 & 0.2876 \\
11000 & 10 	& 1.2987 & 0.2655 & 0.5556 & 1.5162 & 0.4306 & 0.2729 \\
12000 & 10 	& 1.2201 & 0.2442 & 0.5092 & 1.3767 & 0.3904 & 0.2601 \\
13000 & 10 	& 1.1513 & 0.2259 & 0.4695 & 1.2589 & 0.3566 & 0.2488 \\
14000 & 10 	& 1.0906 & 0.2100 & 0.4352 & 1.1582 & 0.3277 & 0.2389 \\
15000 & 10 	& 1.0365 & 0.1960 & 0.4052 & 1.0712 & 0.3028 & 0.2299 \\
16000 & 10 	& 0.9880 & 0.1837 & 0.3788 & 0.9954 & 0.2810 & 0.2219 \\
17000 & 10 	& 0.9442 & 0.1727 & 0.3554 & 0.9287 & 0.2620 & 0.2146 \\
18000 & 10 	& 0.9044 & 0.1629 & 0.3345 & 0.8697 & 0.2451 & 0.2079 \\
19000 & 10 	& 0.8680 & 0.1540 & 0.3157 & 0.8172 & 0.2301 & 0.2017 \\
20000 & 10 	& 0.8347 & 0.1460 & 0.2988 & 0.7701 & 0.2166 & 0.1961 \\
10000 & 100 	& 1.4005 & 0.2910 & 0.6116 & 1.6872 & 0.4796 & 0.2978 \\
11000 & 100 	& 1.3115 & 0.2661 & 0.5571 & 1.5240 & 0.4326 & 0.2850 \\
12000 & 100 	& 1.2349 & 0.2449 & 0.5113 & 1.3889 & 0.3938 & 0.2741 \\
13000 & 100 	& 1.1681 & 0.2268 & 0.4722 & 1.2755 & 0.3614 & 0.2644 \\
14000 & 100 	& 1.1092 & 0.2111 & 0.4385 & 1.1792 & 0.3338 & 0.2559 \\
15000 & 100 	& 1.0568 & 0.1973 & 0.4091 & 1.0964 & 0.3102 & 0.2482 \\
16000 & 100 	& 1.0098 & 0.1851 & 0.3833 & 1.0245 & 0.2898 & 0.2411 \\
17000 & 100 	& 0.9673 & 0.1743 & 0.3604 & 0.9616 & 0.2720 & 0.2347 \\
18000 & 100 	& 0.9286 & 0.1647 & 0.3401 & 0.9061 & 0.2563 & 0.2287 \\
19000 & 100 	& 0.8933 & 0.1560 & 0.3218 & 0.8571 & 0.2424 & 0.2233 \\
20000 & 100 	& 0.8609 & 0.1481 & 0.3054 & 0.8133 & 0.2300 & 0.2183 \\
10000 & 1000 	& 1.4732 & 0.2939 & 0.6206 & 1.7530 & 0.4969 & 0.3759 \\
11000 & 1000 	& 1.4004 & 0.2700 & 0.5698 & 1.6164 & 0.4576 & 0.3775 \\
12000 & 1000 	& 1.3393 & 0.2501 & 0.5279 & 1.5090 & 0.4269 & 0.3793 \\
13000 & 1000 	& 1.2868 & 0.2333 & 0.4930 & 1.4233 & 0.4027 & 0.3808 \\
14000 & 1000 	& 1.2408 & 0.2189 & 0.4635 & 1.3540 & 0.3835 & 0.3815 \\
15000 & 1000 	& 1.1998 & 0.2064 & 0.4382 & 1.2970 & 0.3680 & 0.3814 \\
16000 & 1000 	& 1.1627 & 0.1955 & 0.4164 & 1.2493 & 0.3554 & 0.3804 \\
17000 & 1000 	& 1.1285 & 0.1859 & 0.3973 & 1.2086 & 0.3448 & 0.3785 \\
18000 & 1000 	& 1.0969 & 0.1775 & 0.3805 & 1.1740 & 0.3359 & 0.3762 \\
19000 & 1000 	& 1.0678 & 0.1700 & 0.3659 & 1.1457 & 0.3284 & 0.3745 \\
20000 & 1000 	& 1.0405 & 0.1632 & 0.3528 & 1.1206 & 0.3217 & 0.3721 \\

\hline
\end{tabular}
\label{table:emis}
\end{table*}

\bsp

\label{lastpage}
\clearpage
\end{document}